\documentclass[aps,pra,amsmath,amssymb,superscriptaddress,floatfix]{revtex4-2}%

\usepackage{xspace} 
\usepackage{graphicx}
\usepackage{bm}
\usepackage{xcolor}
\usepackage{color}

\begin{document}

\title{Effect of Proton Irradiation in Thin-Film YBa$_2$Cu$_3$O$_{7-\delta}$ Superconductor}

\author{Joseph Fogt}
\affiliation{Department of Physics, Hope College, Holland, MI 49423, USA}
\email{cho@hope.edu}

\author{Hope Weeda}
\affiliation{Department of Physics, Hope College, Holland, MI 49423, USA}

\author{Trevor Harrison}
\affiliation{Department of Physics, Hope College, Holland, MI 49423, USA}

\author{Nolan Miles}
\affiliation{Department of Physics, Hope College, Holland, MI 49423, USA}

\author{Kyuil Cho}
\affiliation{Department of Physics, Hope College, Holland, MI 49423, USA}

\date{\today}

\begin{abstract}
We investigated the effect of 0.6 MeV proton irradiation on the superconducting and normal state properties of thin-film $\text{YBa}_{2}\text{Cu}_{3}\text{O}_{7-\delta}$ superconductors. A thin-film YBCO superconductor ($\approx$ 567 nm thick) was subject to a series of proton irradiations with a total fluence of $7.6\times10^{16}$ $\text{p/cm}^2$. Upon irradiation, $T_c$ was drastically decreased from 89.3 K towards zero with a corresponding increase in its normal state resistivity above $T_c$. This increase in resistivity which indicates the increase of defects inside the thin-film sample can be converted to the dimensionless scattering rate. We found that the relation between $T_c$ and dimensionless scattering rate obtained during proton irradiation approximates the generalized d-wave Abrikosov-Gor'kov theory better than the previous results obtained from electron irradiations. This is an unexpected result since the electron irradiation is known to be most effective to suppress superconductivity over other heavier ion irradiations such as proton irradiation. It suggests that the type of defects created by proton irradiation evolves from cascade defects (in bulk single crystals) to point-like defects (in thin-film single crystals) as the thickness decreases.
\end{abstract}
\maketitle

\section{Introduction}

$\text{YBa}_{2}\text{Cu}_{3}\text{O}_{7-\delta}$ (YBCO) is one of the most heavily studied superconductors due to its high critical temperature~\cite{Bednorz1986, Wu1987} and practical applications such as superconducting magnets~\cite{Hahn2019Nature_HTS-magnet} and nuclear fusion reactors~\cite{Molodyk2021SciRep_HTS-fusion}. While its Cooper pairing mechanism is not entirely understood, it is accepted that optimally doped YBCO ($T_c$ $\approx$ 93 K) has nodal d-wave pairing symmetry of the order parameter~\cite{Xu1995, vanHarlingenRMP1995, Annett1996, SHEN200814, Tsuei2000RMP_cuprate_review}. 

One useful method for determining the pairing symmetry of the order parameter is to introduce disorders into the crystal structure of a superconductor and investigate the response of its superconducting properties. Depending on the pairing symmetry, disorders will have different effects on the superconducting properties. In an isotropic s-wave superconductor, non-magnetic disorders are not effective in suppressing superconductivity (so-called Anderson's theorem)~\cite{Anderson1959a}. However, magnetic disorders are effective scatterers to suppress $T_c$ of s-wave superconductors (Abrikosov-Gor'kov theory or AG theory)~\cite{AbrikosovGorkov1960ZETF}. In an anisotropic d-wave superconductor, non-magnetic disorders are also effective scatterers to suppress superconductivity~\cite{Radtke1993}. Therefore, Openov {\it{et al.}} developed a generalized AG theory that can include the effect of non-magnetic disorders on the property of d-wave superconductors~\cite{Openov1997}.

The effect of disorders on the properties of superconductors has been experimentally investigated by conducting high-energy particle irradiations to create artificial defects in various superconductors. Different types of high-energy particles have been used such as electrons~\cite{Rullier-Albenque2003PRL_YBCO_e-irr, Giapintzakis1995, Cho2016SciAdv_BaK122, Cho2022PRB_YBCO_e-irr}, protons~\cite{Wu1993PRB_YBCO_thin-film, Torsello2022IEEE_FeSe_thinfilm_proton-irr}, and heavy ions~\cite{KonczykowskiPRB1991_HeavyIon_YBCO, Nakajima2009PRB_HeavyIon}. Electron irradiation is known to generate atomic-size point defects (due to its low rest mass)~\cite{Giapintzakis1995}; proton irradiation to generate a cascade of point defects; and heavy-ion irradiation to generate columnar defects. Among them, electron irradiation which can create point-defects is known to be most effective in suppressing superconductivity. Indeed, electron irradiation studies performed on YBCO compounds show qualitative agreement with the generalized d-wave AG theory. However, the results are still far from quantitative agreement. Therefore, this disagreement was explained in diverse ways such as the quality of each sample using different plasma frequencies~\cite{Radtke1993, Openov1998}, the ratio between in-plane and out-of-plane defects~\cite{GraserHirschfeld2007PRB_cuprates_defects}, and the strong correlation~\cite{Garg2008NaturePhysics, KemperHirschfeld2009PRB, Tang2016PRB_strong_correlation_d-wave}.

\begin{figure}
    \centering
    \includegraphics[width=8.5cm]{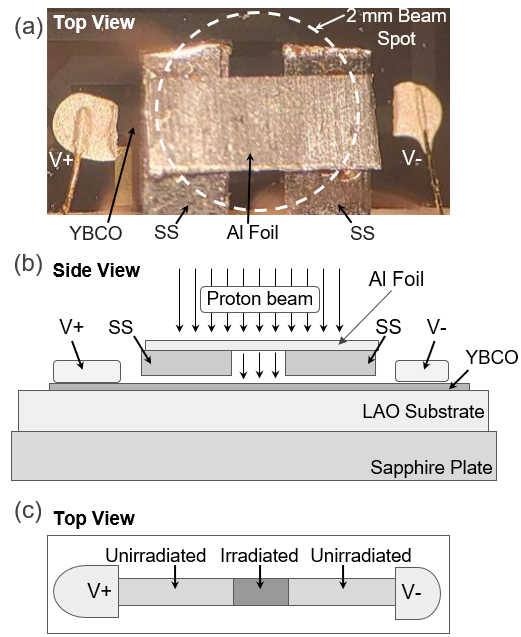}
    \caption{YBCO thin-film sample patterned on LAO substrate. It is prepared for resistance measurement and proton irradiation. (a) Top-view that shows the section where the resistance is measured between V+ and V-. I+ and I- are located outside of this image. The white dashed circle indicates a 2 mm diameter spot of the proton beam. 50 $\mu$m thick aluminum foil (energy degrader) was placed above the sample to decrease the proton beam energy from 2.2 MeV to 0.6 MeV. The degrader was placed right above two stainless-steel (SS) plates (0.19 mm thick). Since the SS plates completely block the proton beam, only the center area of the YBCO sample between the SS plates was irradiated. (b) Side-view that shows the aluminum energy degrader and two SS blocks placed above the YBCO thin-film without physical contact. The YBCO thin-film was epitaxially grown on the LAO substrate. YBCO/LAO was silver-pasted on a sapphire plate. (c) Schematic diagram that shows the areas of irradiated and unirradiated parts of the YBCO thin-film. Proton irradiation was applied only to the center part of the thin-film.}
    \label{fig:target setup}
\end{figure} 

To understand the relation between disorders and $T_c$ of d-wave superconductors, we conducted 0.6 MeV proton irradiation in thin-film YBCO superconductor ($\approx$ 567 nm thick) using Hope College’s particle accelerator (1.7 MV tandem Van de Graaf electrostatic accelerator). After a series of irradiation and resistance measurements, we obtained the relation between $T_c$ and normal state resistivity. After converting normal state resistivity to dimensionless scattering rate, we compared our result with the generalized d-wave AG theory and previous electron irradiation results. It was found that our results agree better with the generalized d-wave AG theory than previous electron irradiation studies. This is an unexpected outcome since the electron irradiation is known to be more effective in suppressing superconductivity than other forms of irradiation such as proton irradiation. This unexpected result suggests that the type of defects created by proton irradiation evolves from cascade defects in bulk compounds to point-like defects in thin-film compounds as the thickness decreases.

\section{Materials and Methods}

\subsection{YBCO thin-film single crystal}
The YBCO thin-film ($\approx$ 567 nm thick) was epitaxially grown on a lanthanum aluminate (LaAlO$_3$, or LAO) substrate. Photoresist was spin-coated onto the film, baked, exposed under a mask to UV light, and milled with an Ar ion beam. After patterning, the film was annealed in O$_2$ at 500 $^{\circ}$C for one hour. The sample was originally fabricated as resonators in commercial microwave filters for wireless base stations~\cite{Remillard2014SST_YBCO}. The sample shows $T_{c}$ $\approx$ 89.3 K indicating that its superconducting property is close to the bulk single crystalline sample of $T_c$ $\approx$ 93 K.

\subsection{Resistance measurement}
\begin{figure}
    \centering
    \includegraphics[width=8.5cm]{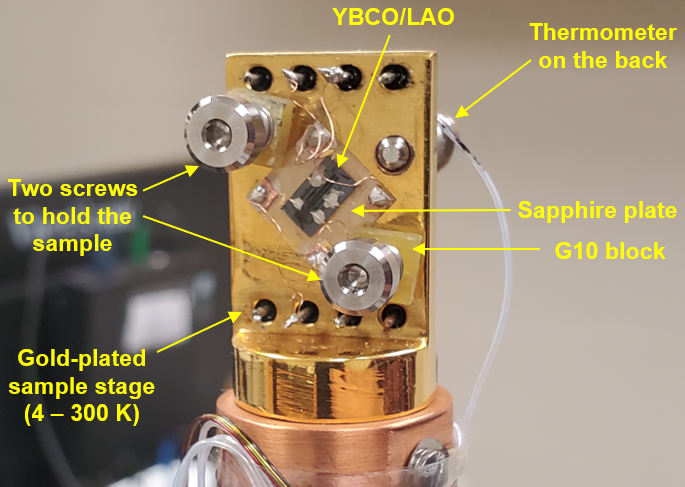}
    \caption{YBCO/LAO sample mounted on a 4K cryostat. Two screws hold the sapphire plate on which the YBCO/LAO sample is silver-pasted.}
    \label{fig:cryostat setup}
\end{figure}

The in-plane resistance of the YBCO thin-film was measured using a standard four-probe technique. Fig.~\ref{fig:target setup} (a) shows the part of the sample where the resistance is measured between V+ and V-. I+ and I- are located outside of the image. The dimensions of the measured part of the sample are 2.663 ($\pm$ 0.016) mm $\times$ 0.2570 ($\pm$ 0.0008) mm $\times$ 566.7 ($\pm$ 1.9) nm. Four electrical contacts made of thin gold wires were adhered to the thin-film using silver-paste. The YBCO/LAO sample was attached to a sapphire plate using silver-paste and the whole setup was mounted on the gold-plated sample stage of a 4K cryostat with two screws (Fig.~\ref{fig:cryostat setup}) for temperature-dependent resistance measurements. 

\subsection{Energy degrader}

The Hope College's particle accelerator can directly generate a proton beam of energy ranging from 0.6 MeV to 3.4 MeV. However, the low-energy operation at 0.6 MeV is unsafe for a long period of operation since the low-energy beam can unintentionally damage the beam line. Therefore, we developed an alternative way to generate a 0.6 MeV proton beam. We first generated a 2.2 MeV proton beam and then passed this beam through an aluminum energy degrader to decrease the beam energy from 2.2 MeV to 0.6 MeV. 

\begin{figure}
    \centering
    \includegraphics[width=8 cm]{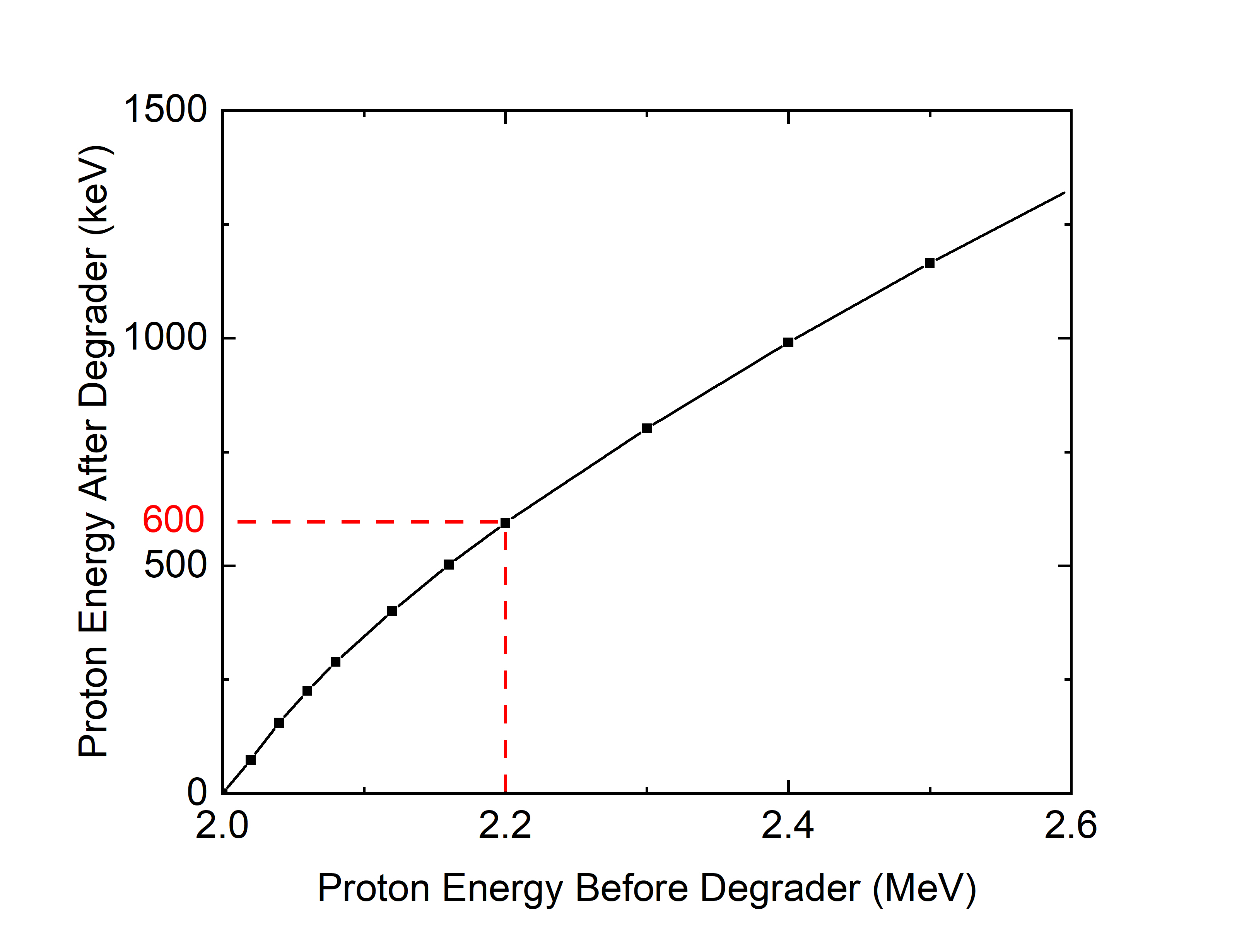}
    \caption{Proton energy before and after an aluminum energy degrader calculated using SRIM software. The red dotted line indicates the energy used for the current study (about 600 keV).}
    \label{degradation graph}
\end{figure}

To accomplish this, we conducted a series of proton irradiations directly onto an aluminum degrader (50 $\mu$m foil measured by a micrometer) by varying energy from 1.9 to 2.6 MeV and measured the beam current after the beam passed through the aluminum degrader using a Faraday cup. It was found that the beam current after the degrader became zero when the incident beam energy before the degrader was lower than 2.0 MeV. This suggests that the penetration distance of the proton beam is smaller than the thickness of the aluminum degrader when the accelerator energy is lower than 2.0 MeV. Using SRIM software~\cite{Zieglera2010SRIM}, it is found that the effective thickness of the aluminum degrader corresponding to 2.0 MeV is 41.63 $\mu$m (smaller than 50 $\mu$m measured by the micrometer).

The effective thickness of the degrader (41.63 $\mu$m) was used to calculate the beam energies after degradation by the aluminum degrader as shown in Fig.~\ref{degradation graph}. For example, the projected penetration distance for a 2.2 MeV proton into an infinitely thick aluminum degrader is 48.49 $\mu$m according to SRIM. Since the effective thickness of the aluminum degrader is 41.63 $\mu$m, the beam still has some energy left after traveling through the target. Thus, the difference between 48.49 $\mu$m and 41.63 $\mu$m (i.e., 6.86 $\mu$m) is the distance that a proton may travel after interacting with the aluminum target. This value can be converted to the energy after degradation which is 0.6 MeV in the case of a 2.2 MeV beam. This process was repeated for multiple beam energies and the resulting data was plotted in Fig.~\ref{degradation graph}.

\subsection{Homogeneous proton beam}

\begin{figure}
    \centering
    \includegraphics[width=8 cm]{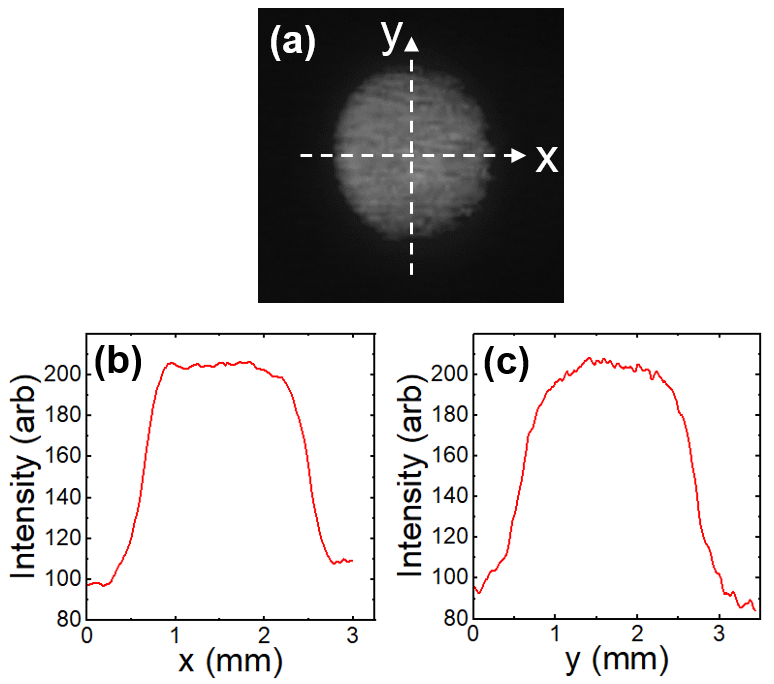}
    \caption{(a) 2 mm diameter proton beam projected on Mylar scintillator. (b) and (c): Intensity of the beam along the x and y axes.}
    \label{fig:beam profile}
\end{figure}

The 0.6 MeV proton irradiation was conducted at room temperature in a vacuum chamber of $6 \times 10^{-8}$ torr. The beam was directed along the c-axis of the YBCO thin-film sample. We used a 2 mm diameter aperture to generate a 2 mm diameter proton beam. The beam profile shown in Fig.~\ref{fig:beam profile} indicates that its center area is homogeneous. The actual positions of YBCO sample, the 2 mm beam spot, and the aluminum energy degrader are described in Fig.~\ref{fig:target setup} (a). In addition, two 0.19 mm-thick stainless-steel (SS) plates were placed above the sample without physical contact. Since the proton beam cannot penetrate the SS plates, the irradiation only affected the area between two SS plates. The irradiated part of the YBCO thin-film (0.591 ($\pm$ 0.013) mm $\times$ 0.257 ($\pm$ 0.0008) mm $\times$ 566.7 ($\pm$ 1.9) nm), which is much smaller than 2 mm diameter beam spot, was positioned at the center of the beam to experience the most homogeneous beam (Fig.~\ref{fig:target setup}). Figure~\ref{fig:target setup} (c) is a schematic diagram that shows where the irradiation was performed. In this way, we generated a homogeneous 0.6 MeV proton beam on the well-defined section of the YBCO thin-film. The current of the beam was measured using a Faraday cup every 30 minutes during the irradiation. The current of the 2 mm diameter beam remained stable at 10 nA ($\pm$ 2 nA) during the whole irradiation. To avoid sample heating, the current was kept below 12 nA. After irradiation, the sample was transferred from the accelerator vacuum chamber to the cryostat (Fig.~\ref{fig:cryostat setup}) for resistance measurement. 

\section{Results and Discussion}

\begin{figure}
\centering
\includegraphics[width=13.5 cm]{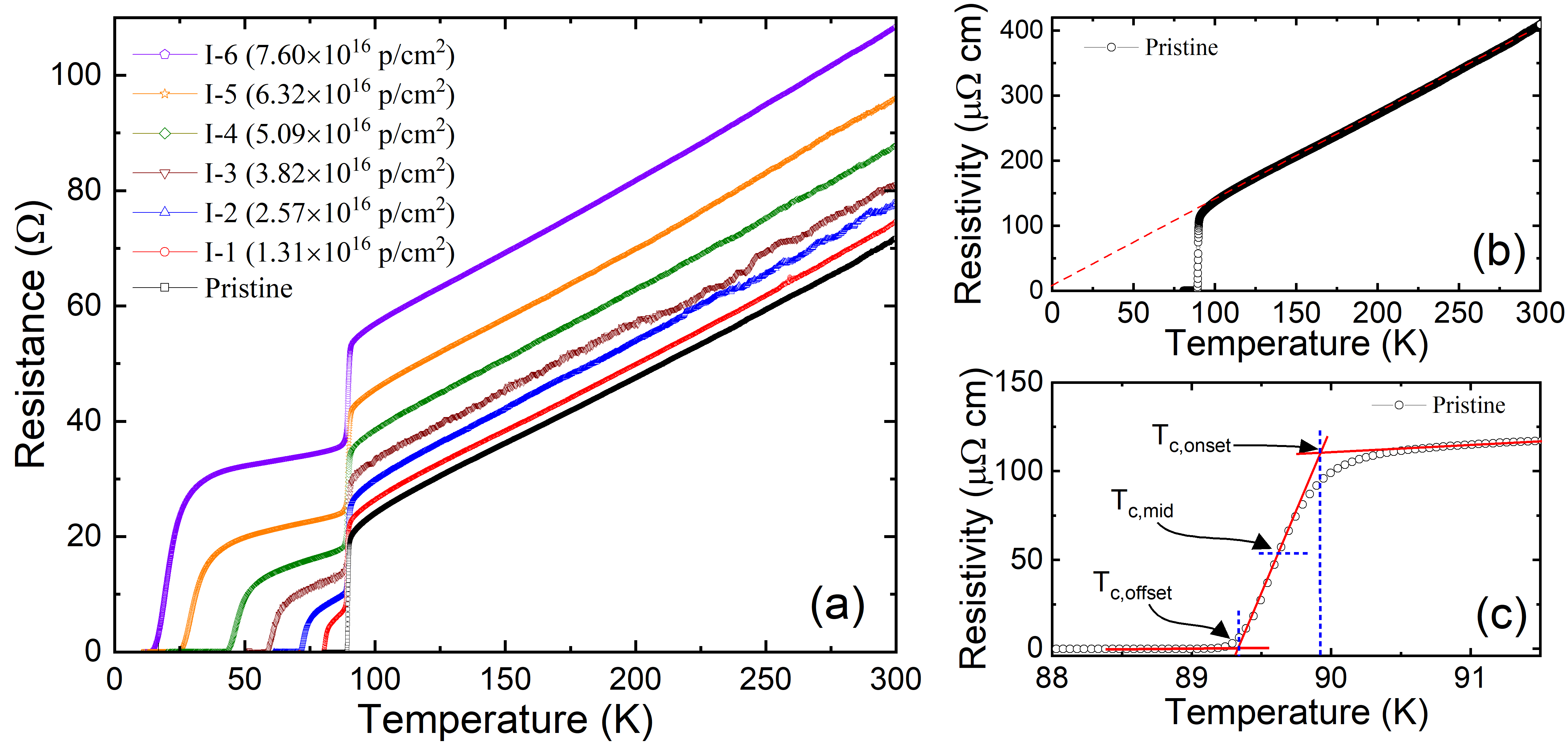 } 
\caption{(a) Temperature-dependent resistance measured in YBCO thin-film sample upon a series of irradiation (from I-1 to I-6). (b) Temperature-dependent resistivity of the pristine sample. (c) Zoom-in of the panel (b) near the superconducting transition. Among different definitions of $T_c$, we used the offset definition ($T_{c,offset}$) throughout this article.} 
\label{fig:Resistance Scaled}
\end{figure}

\begin{figure}
\centering
\includegraphics[width= 8 cm]{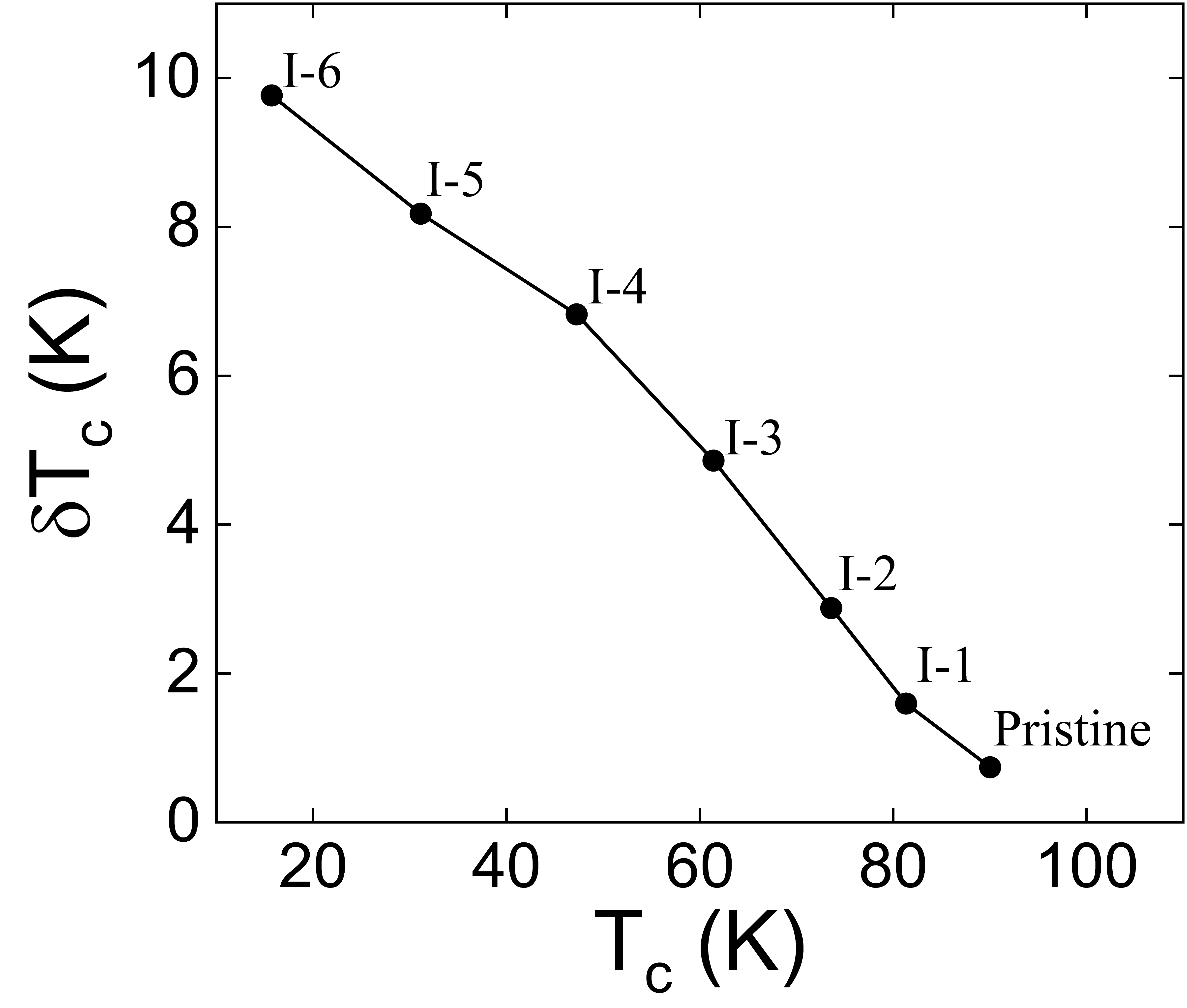} 
\caption{$\delta T_c$ = $T_{c,onset} - T_{c,offset}$ that shows broadening of superconducting transition. It increases upon proton irradiation.}
\label{fig:Tc_broadening}
\end{figure}

We performed seven resistance measurements alternated with six proton irradiations in the identical YBCO sample. Figure~\ref{fig:Resistance Scaled} (a) summarizes all resistance measurements. Except for the pristine case, all the other data show two superconducting transitions. Two superconducting transitions are observed since the irradiation was only performed on the center part of the YBCO sample as shown in Fig.~\ref{fig:target setup}(c). Therefore, $T_c$ of the irradiated part decreases upon irradiation while $T_c$ of the unirradiated part remains unchanged. Two-step transitions are not seen for the pristine case because its entire area has not been irradiated. In the irradiated portion of the sample, the damage to superconducting properties is evident with $T_{c}$ decreasing rapidly from the initial $T_c$ of 89.3 K towards zero. Furthermore, the transitions get broader as the fluence increases as shown in Fig.~\ref{fig:Tc_broadening}. Figure~\ref{fig:Resistance Scaled} (b) is the temperature-dependent resistivity of the pristine sample calculated based on the shape of the sample. The linear approximation of the normal state resistivity suggests very small residual resistivity ($\approx$ 8 $\mu \Omega cm$) at $T$ = 0 K. Figure~\ref{fig:Resistance Scaled} (c) shows $T_{c,offset}$ that is used for data analysis. 

Figure~\ref{fig:Resistance Scaled} (a) also shows that upon proton irradiation the normal state resistance above $T_c$ monotonically increases over the entire high-temperature region. This parallel upward shift in resistance is consistent with Matthiessen’s rule, suggesting that the number of defects in the YBCO sample gradually increases upon irradiation. In addition, this increase in resistance originates from the irradiated part of the sample, while the resistance in the unirradiated part remains unchanged. Using the exact volume of the irradiated part of the sample, the resistance increase can be converted to the resistivity increase ($\Delta \rho$) of the irradiated part. Since the normal state resistance increase is almost identical in all temperature regions above $T_c$, the resistance value at 125 K was used to represent the increase of defects and to calculate the resistivity increase at 125 K ($\Delta \rho_{125 K}$). 

The generalized AG theory formulated by Openov~\cite{Openov1998} can be written for the case of the non-magnetic disorders in d-wave superconductors as follows,

\begin{equation}
    -ln(t_c ) = \Psi ( \frac{1}{2} + \frac{g}{2~t_c} ) - \Psi (\frac{1}{2}),
    \label{eqn:Generalized_AG_theory}
\end{equation}
where $t_c$ is $T_c$/$T_{c0}$, $T_{c0}$ is the initial $T_c$ before the disorders are added, and $g$ is the dimensionless scattering rate. $t_c$ asymptotically goes to zero as $g$ approaches to 0.28. Using the Drude model~\cite{Prozorov2014, Cho2018SST_review}, $g$ can also be written in terms of the residual resistivity ($\rho_{0}$) as follows, 

\begin{equation}
    g =\frac{\hbar~\rho_{0}}{2\pi k_B \mu_0 T_{c0} \lambda_0^2},
    \label{eqn:Dimensionless_Scattering}
\end{equation}
where $\rho_{0}$ is the residual resistivity at $T$ = 0 K of the irradiated part of the YBCO sample, $T_{c0}$ is the critical temperature of the pristine sample, and $\lambda_0$ is the zero-temperature London penetration depth of the pristine sample. Due to the high $T_c$ of the YBCO sample, it is difficult to estimate the exact residual resistivity. However, the temperature-dependent resistivity in Fig.~\ref{fig:Resistance Scaled}(b) shows that the linear approximation of normal state resistivity suggests very small residual resistivity at $T$ = 0 K ($\approx$ 8 $\mu \Omega cm$). Assuming that $\rho_{0}$ of the pristine sample is very small, $\rho_{0}$ in Eq.~\ref{eqn:Dimensionless_Scattering} is replaced with $\Delta \rho$ and $g$ can be rewritten as follows,

\begin{equation}
    g \approx \frac{\hbar~\Delta\rho_{125 K}}{2\pi k_B \mu_0 T_{c0} \lambda_0^2},
    \label{eqn:Dimensionless_Scattering_2}
\end{equation}
where $\Delta\rho_{125 K}$ is the resistivity increase at $T$ = 125 K of the irradiated part of the YBCO sample. Since $\lambda_0$ varies in different studies such as $1460\pm150$ \r{A} \cite{Prozorov2000APL}, 1550 \r{A} \cite{Tallon1995}, and $1990\pm200$ \r{A} \cite{Djordjevic1998}, we used a most commonly accepted value of $1405\pm92$ \r{A}~\cite{Bonn1933}. 

\begin{figure}
\centering
\includegraphics[width=10cm]{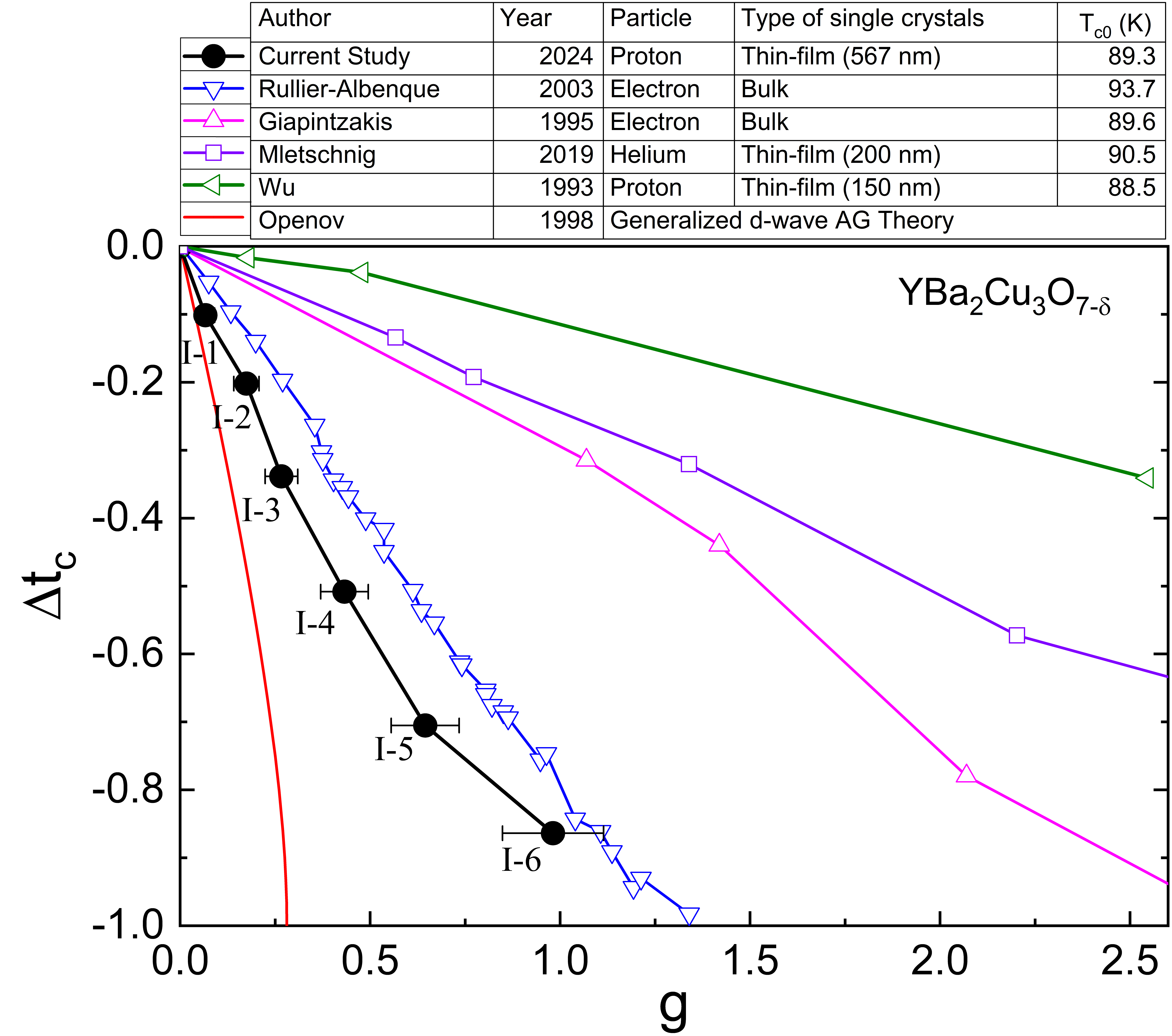} 
\caption{$\Delta t_c$ = ($T_c - T_{c0}$)/$T_{c0}$ as a function of dimensionless scattering rate ($g$) upon irradiation. The current proton irradiation result is compared with previous results~\cite{Rullier-Albenque2003PRL_YBCO_e-irr, Giapintzakis1995, Mletschnig2019, Wu1993PRB_YBCO_thin-film} and the generalized d-wave AG theory~\cite{Openov1998}.}
\label{fig:Dimensionless Scattering}
\end{figure}

Figure~\ref{fig:Dimensionless Scattering} summarizes the relation between $T_c$ and $g$ of the current study in comparison with previous irradiation studies~\cite{Giapintzakis1995, Rullier-Albenque2003PRL_YBCO_e-irr, Mletschnig2019, Wu1993PRB_YBCO_thin-film} and theoretical expectation~\cite{Openov1998}. The current proton irradiation study shows that $T_c$ linearly decreases down to $\Delta t_c$ $\approx$ -0.7 (I-5) with increasing $g$. After I-5 irradiation, however, the decrease rate slows down for I-6. Comparing the current study with previous irradiation studies on YBCO single-crystals (bulk sample and thin-film), it is evident that the current study of proton irradiation on YBCO thin-film are closest to the theoretical expectation (generalized d-wave AG theory by Openov~\cite{Openov1998}). It is interesting to find that the current results show a faster suppression rate of $T_c$ than that of electron irradiation study performed on a bulk single-crystalline YBCO by Rullier-Albenque {\it{et al.}}~\cite{Rullier-Albenque2003PRL_YBCO_e-irr}.
This outcome is unexpected since the electron irradiation has been commonly known to be most effective in suppressing $T_c$ by producing atomic-size point-like defects, while other heavier ion irradiations produce less effective cascade or columnar defects. Therefore, the current proton irradiation results on a thin-film YBCO superconductor follow the generalized d-wave AG theory better than the electron irradiation results. This unexpected outcome might be attributed to the use of thin-film single-crystalline samples, while the other electron irradiation studies were performed on large bulk single-crystalline samples. For thick samples, the proton beam creates cascade defects deep inside the sample. However, as the thickness of the sample is reduced to the thin-film limit, the dominant defects varies from cascade defects to atomic-size point-defects. Since the sample thickness in the current study ($\approx$ 567 nm) is much smaller than the implantation depth (3.9 $\mu$m) of 0.6 MeV proton beam, the atomic-size point-defects is expected to dominate. 

\begin{figure}
\centering
\includegraphics[width=8.5cm]{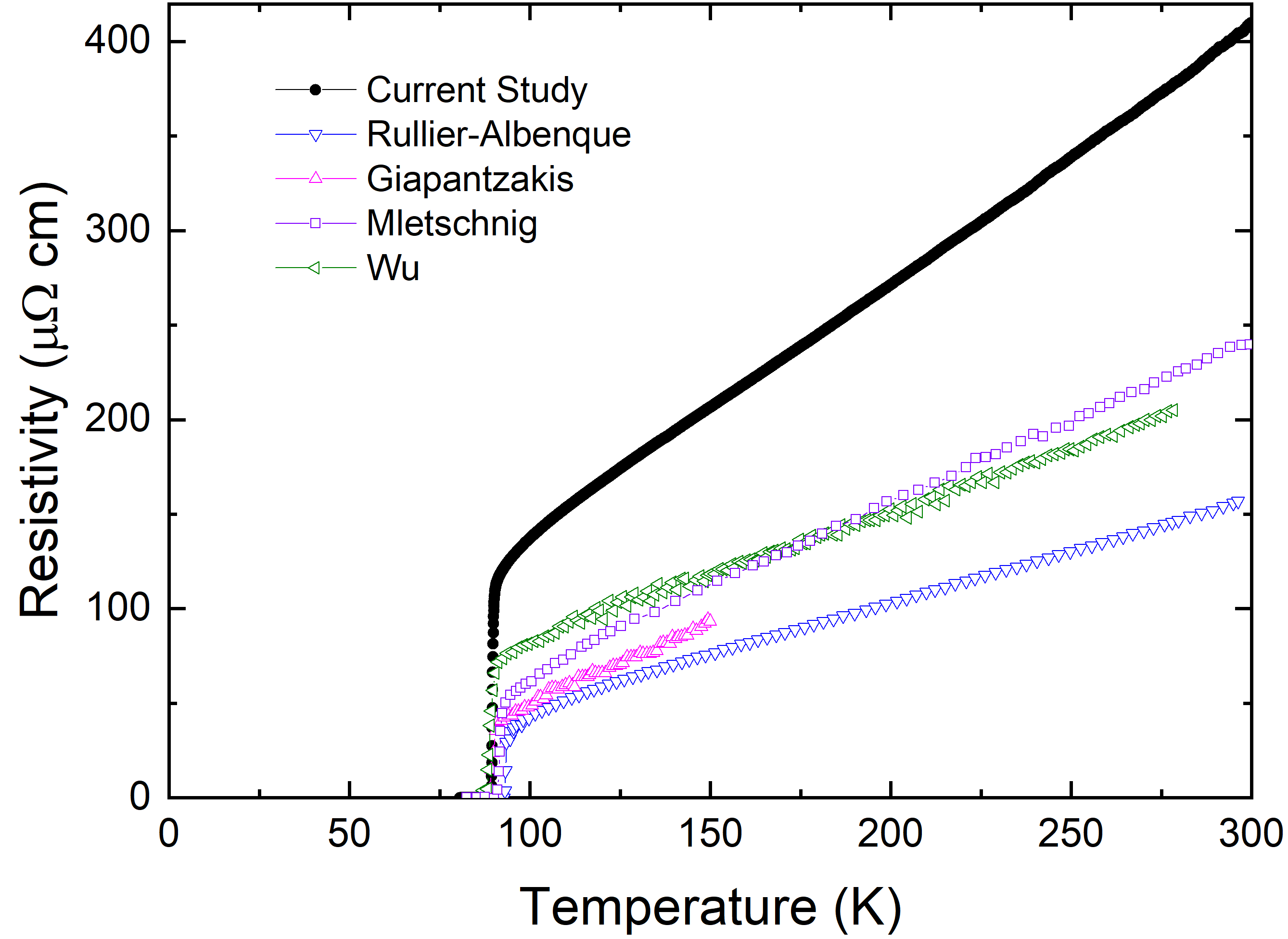} 
\caption{Comparison of temperature-dependent resistivities measured in different YBCO single crystals~\cite{Rullier-Albenque2003PRL_YBCO_e-irr, Giapintzakis1995, Mletschnig2019, Wu1993PRB_YBCO_thin-film}. The resistivities of thin-film samples are commonly larger than those of bulk single crystalline samples.}
\label{fig:resistivity_comparison}
\end{figure}

A noticeable departure from this trend occurs for I-6 irradiation. $T_c$ decreases at a slower rate for I-6 than the others (from I-1 to I-5). This indicates a development of extended defects where the nearby point defects agglomerates at high fluence as mentioned by Wu {\it{et al.}}~\cite{Wu1993PRB_YBCO_thin-film}. In addition, we noticed a height increase on the area of the irradiated part of the YBCO thin-film by about 15 nm (identified from Atomic Force Microscopy). This increase was caused by implantation of protons into the LAO substrate (implantation depth of protons into YBCO/LAO = 3.9 $\mu$m from TRIM simulation~\cite{Zieglera2010SRIM}). The height increase upon irradiation was also observed by Zhao {\it{et al.}}~\cite{Zhao1991PhysicaC_YBCO_thin-film}. The height increase can cause distortion of the YBCO thin-film by increasing the extra resistivity at the boundary between the irradiated part and unirradiated part of YBCO thin-film. Further investigations are needed to check the influence of height increase on the suppression rate of $T_c$.

Figure~\ref{fig:resistivity_comparison} compares temperature-dependent resistivities of various YBCO samples before particle irradiations. Thin-film single-crystalline samples show higher resistivity than bulk single-crystalline samples. In particular, the sample of the current study shows higher resistivity and lower $T_c$ than those of the previous studies, suggesting that thin-films may already have strong pinning defects before irradiations. Further investigations are needed to study the effect of proton irradiation by varying initial conditions of the sample such as different $T_{c0}$, resistivity, and thickness. 

\section{Conclusions}

We studied the effect of 0.6 MeV proton irradiation in a YBCO thin-film superconductor and found that the proton irradiation on thin-film limit YBCO crystal can produce atomic-size point-like defects. Indeed, the relation between $T_c$ and the dimensionless scattering rate $g$ approximated the generalized d-wave Abrikosov-Gor'kov theory, surpassing the results of electron irradiation studies. Further investigations are needed to understand how the type of defects induced by proton irradiations varies as the thickness of the sample varies.

\acknowledgments{We thank Dr. Ruslan Prozorov, Dr. Danielle Torsello, Dr. Peter Hirschfeld and Dr. Vivek Mishra for the helpful discussion. We thank Dr. Paul DeYoung and Mr. Andrew Bunnell for assisting in operating the particle accelerator. This work was supported by the Dean of Natural and Applied Sciences and the Physics Department of Hope College. This research was partially funded by by the National Aeronautics and Space Administration (NASA), under award number 80NSSC20M0124, Michigan Space Grant Consortium (MSGC).}

\bibliographystyle{apsrev4-2}
\bibliography{YBCO_p-irr_arXiv}

\end{document}